# CONTROL IN EPICS FOR CONDITIONING TEST STANDS FOR ESS

A. Gaget[†], A. Gomes, Y. Lussignol, CEA Saclay Irfu DIS, France

*Abstract*

CEA Irfu Saclay is involved as partner in the ESS accelerator construction through different work-packages: controls for several RF test stands, for cryomodule demonstrators, for the RFQ coupler test and for the conditioning around 120 couplers and the tests of 8 cryomodules. Due to the high number of components it is really crucial to automatize the conditioning. This paper describes how the control of these test stands was done using the ESS EPICS Environment and homemade EPICS modules. These custom modules were designed to be as generic as possible for reuse in future similar platforms and developments. They rely on the IOxOS FMC ADC3111 acquisition card, Beckhoff EtherCAT modules and the MRF timing system.

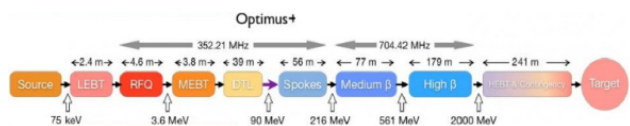

Figure 1: ESS acceleration stages.

## INTRODUCTION

The European Spallation Source ESS is a large European research infrastructure under construction in Lund, Sweden. It will be composed among others of an RFQ and a total of 30 elliptical cryomodules (see Fig. 1). These will be integrated in the next 3 years with a delivery rate of one cryomodule per month. An international collaboration has been established to develop and construct the 30 elliptical cryomodules. CEA Saclay and IPN Orsay are collaborating to design, build and test a first Medium beta Elliptical Cavities Cryomodule Demonstrator (M-ECCTD). A second demonstrator with high beta cavities (H-ECCTD) is being developed by CEA before starting the cryomodules production. CEA will provide many other components, such as the power couplers and their RF processing, it will also provide the power couplers of the RFQ and their processing.

The control for all these activities has been developed in EPICS, based on the ESS EPICS Environment (EEE) [1] developed by the ESS control team at Lund (ICS). This environment integrates the use of versioned EPICS modules and allows genericity. Moreover, it provides EPICS drivers and modules for multiple hardware. Due to the numerous components to condition and several similar test stands, it appeared that an EPICS module platform with a certain level of genericity would be useful. This paper describes the use of the EEE environment and generic tools that have been developed here at CEA for these test stands.

## DESCRIPTION

### Hardware Solution

An IOxOS solution with the VME64X CPU card IFC1210 coupled with the ADC3111 FMC card [2] for the fast acquisition (see Fig. 2), an MRF solution with the VME EVG 230 card coupled with the VME EVR 230RF card for timing [3]., and Beckhoff EtherCAT modules [4] for the slow acquisition have been chosen. All the drivers for this hardware were developed by ICS.

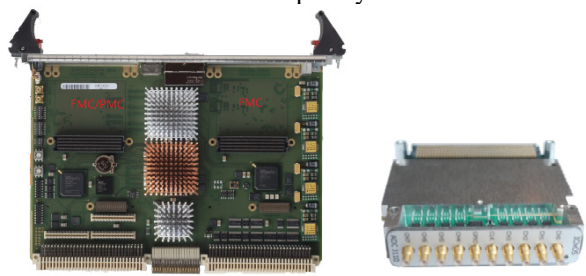

Figure 2: IOxOS cards (IFC1210 and ADC3111).

- The ADC3111 is an FMC mezzanine acquisition board which allows to acquire up to 8 channels at a sampling rate of up to 250 MHz.
- The VME64X CPU IFC1210 is a single board computer on which a real-time Linux kernel runs and allows to plug 2 FMC cards or 1 FMC and 1 PMC as shown in Fig. 3.

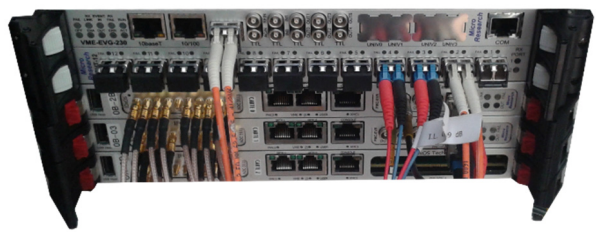

Figure 3: Example of VME rack used.

- MRF Timing system consists of an Event Generator (EVG) which converts timing events and signals to an optical signal distributed through Fan-Out Units to an array of Event Receivers (EVRs) (see Fig. 4). The EVRs decode the optical signal and produce hardware and software output signals based on the timing events received [3].

___

[†] alexis.gaget@cea.fr

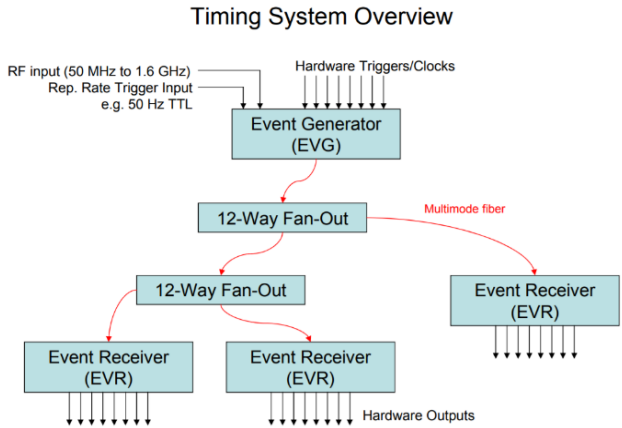

Figure 4: Timing System Overview [5].

*Software Solution*

EPICS modules developed can be distinguished in 3 categories that can use or be used by the other modules.

- Data Modules that produce the raw data from software or from hardware. Mostly drivers from an acquisition hardware (EtherCAT modules, ADC3111…) but it can also be data from other modules that produce data from calculation or from simulation.

- Process Modules that use external data (from data modules) to compute or produce new data depending on its inputs. It takes a waveform or an ai as input with the CPP fields, and so the module will compute a new corresponding output, a waveform or an ao. It means that we can chain or parallelize process modules depending on our needs as shown in Fig. 5.

- Accessories Modules used independently from the other modules but common to a lot of applications (archiving, saveRestore, screenshot...).

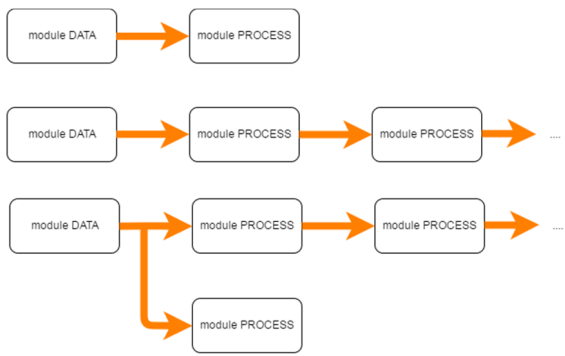

Figure 5: Examples of use case of data and process modules.

Since data modules for ADC3111 and EtherCAT were developed by ICS ESS they won't be detailed in this paper, which will focus on the main process and accessories modules used instead.

**Process modules**

- CalibrationApp is a support application module that converts signal data contained in waveform records using an EPICS breakpoint table or a polynomial function of up to degree 4. It allows different kinds of calibration:
    - Standard breakpoint table calibration. Using the breakpoint table mechanism for a non-linear calibration using a table defined by the developer.

    - RF power calibration: The waveform measurement is converted through a breakpoint table into dBm, then the user can add attenuation to this measurement and it is finally converted into kW with the dBm to Kw formula.

    - RF cavity field calibration: It uses the formula shown in Eq. (1). Each factor except Pt, can be defined by the user dynamically. Pt is computed with the previous method.

$$EAcc = \frac{\sqrt{Qt * \frac{r}{q} * Pt}}{L} \quad (1)$$

Pt = Measurement expressed in mW.
Qt = Cavity coupling factor.
R/Q = Shunt impedance.
L = Cavity length.

    - Polynomial calibration: A polynomial function with a degree up to 4.

- SignalProcessingApp is a support application module that computes statistics on a signal waveform (average, max, min, integral, width…). It manages cursors to delimit a pulse, uses thresholds to delimit the area of interest (see Fig. 6) and creates a compressed signal for GUIs.

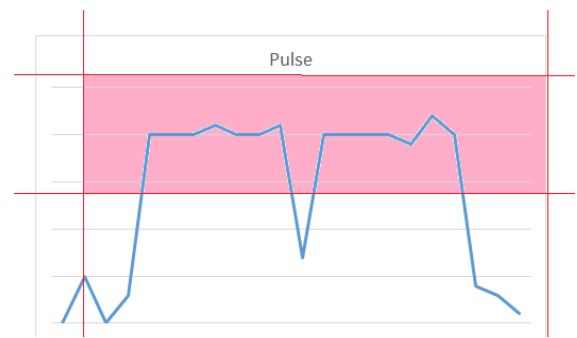

Figure 6: Area where the samples are used for statistics.

- SeqConditioning is a support application module that allows to execute a sequence of RF conditioning. The principle of this sequence is to increase the RF power by incremental steps until a threshold defined by the user, then the sequence is started again with a larger

width pulse (see Fig. 7). It handles 3 types of conditioning default:
- Minor: It pauses the sequence at the actual power until the default has gone.
- Major: It decrements the power until the defaults have gone or until it reaches a limit defined by the user.
- Critical: It stops the sequence.

These 3 defects are just simple Booleans that can be overridden by logical operations. Hence the user adds all the alarms he wishes (although this has to be decided during development, and not still during operation).

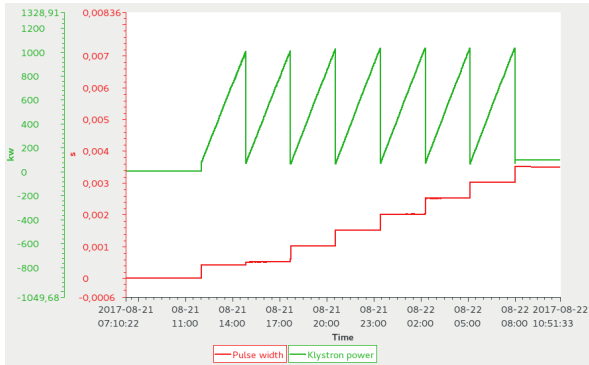

Figure 7: Complete RF Conditioning sequence data.

**Accessories modules**
- PVArchiving is a support application module that allows to pause or resume archiving a PV in an archiver appliance.
- SaveRestore is a support application module that allows to save or restore the value of process variables in files during the execution. It's based on PyEpics save/restore functions [6].

*Network*

The network installation has been installed as shown in Fig. 8. It insures stability and security. The user can access the experiment VLAN or the archiver through a Gateway equipment via another secure VLAN. This gateway filters connection, protocols and data.

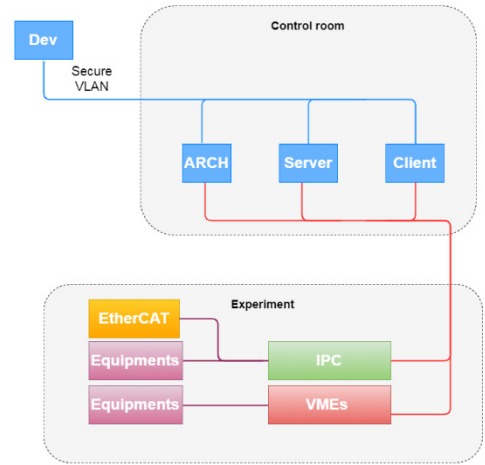

Figure 8: Standard ESS network at CEA Saclay.

## USAGE

The ADC3111 acquires RF signals, photomultiplier (PM) or electron Pick-Up (PUe-). The RF calibration is defined by the user. The PM and PUe- are not calibrated, they are displayed in Volts in the GUI. For each signal, from the ADC, statistics of the signal will be computed (average, max, min, integral, width) and archived.

The EtherCAT modules are used to set thresholds for calibration defects, security interlocks or read measurements (temperature PT100, vacuum…).

The conditioning sequence controls the RF power through an attenuator controlled by an EtherCAT module. It uses vacuum defect (EtherCAT) and information about the RF power (ADC3111) to determine the defect and its behaviour.

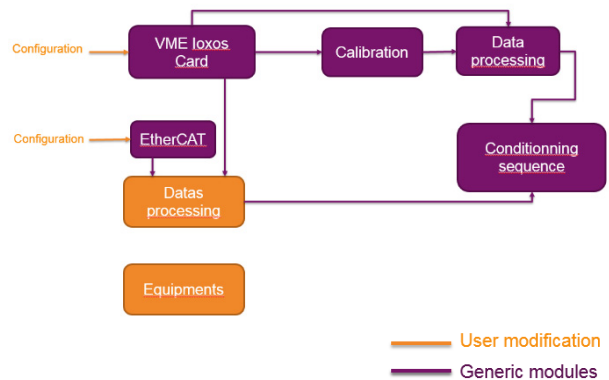

Figure 9: Typical use case.

For each test stand the developer will have to adapt the following details (see Fig. 9):
- The configuration of the IOxOS cards (number, names, attenuation…).
- The configuration of the modules SignalProcessing and CalibrationApp.

- The configuration of EtherCAT modules (number, position, types).
- Data processing if it needs a particular process on data.
- Control of equipment.

So far this has been deployed on 2 sources and 3 test stands.

*Platform RF704 MHz*

**Source 704 MHz** It was originally developed in LabVIEW but has been replaced by an EPICS GUI and Control. In addition to the typical setup described above it also controls security interlocks, via an electronic tray controlled with Modbus, Synthesizer Keysight N5171B and a frequency count Rohde & Schwarz, HM8123 with Ethernet StreamDevice.

**The RF conditioning for couplers cavities 704 MHz.** In additional to the typical setup described above it controls the SNL conditioning sequence.

**The RF conditioning for demonstrator M-ECCTD 704 MHz.** This test stand shares the source 704 MHz with the previous one. In addition to the typical setup described above it controls the piezo motor for adjusting the cavity nominal frequency, with Ethernet StreamDevice**.**

This platform was set up in late 2016. It has already conditioned 8 couplers, used for the demonstrator M-ECCTD among others.

The network has been configured as shown in Fig. 10.

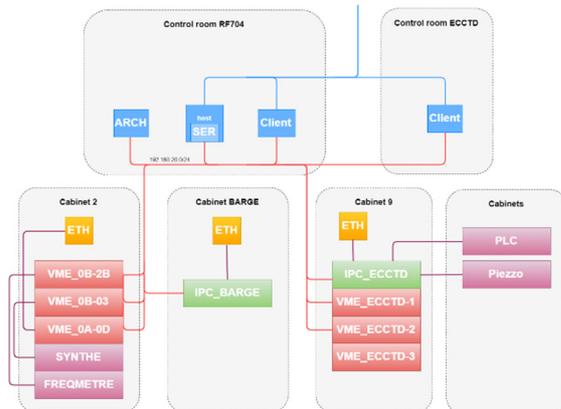

Figure 10: Network scheme of the Platform 704 MHz.

*Platform Pulsed RF352 MHz*

**The source 352 MHz.** In addition to the typical setup described above it controls the DTI Modulator with Modbus. And the Synthesizer Keysight N5171B with Ethernet StreamDevice.

**The RF conditioning for RFQ couplers 352 MHz**. In addition to the typical setup described above it has the control of the SNL conditioning sequence.

The network has been configured as shown in Fig. 11.

The control of the source and the RF process has been tested mid-2017. The RF processing on the RFQ couplers should begin late 2017.

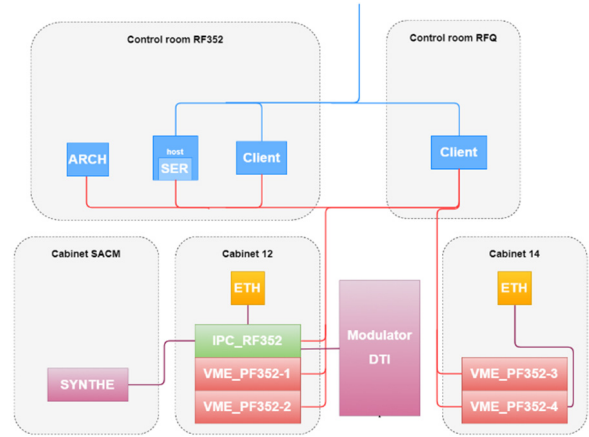

Figure 11: Network scheme of the platform 352 MHz.

## FUTURE APPLICATIONS

*ESS*

These developments have immediate new applications for ESS, since a new RF platform will be installed in Saclay before the end of 2017 to continue conditioning all the elliptical Cryomodules RF power couplers production.

*SARAF-Linac*

CEA-Saclay Irfu is strongly involved in the control of the SARAF-Linac accelerator and has developed its own EPICS environment [7] which includes the generic modules developed for ESS. Therefore this recent development will also be used for SARAF-Linac test stands.

## CONCLUSION

While the software development for the first test stand has certainly been more time consuming because of the genericity constraint, it has been much easier and faster to develop the following two benches. It now remains to validate the different configurations applied to each bench. Besides having been used for the planned benches, every module which was implemented to be self-sufficient may now be reused on different experiments. Some of the "process" modules are already used in the software for the source of ESS in Catania. A module which will be used by more users will benefit from more maintenance and therefore stability.

## ACKNOWLEDGEMENT

The author would like to thank all the people that contributed to the installation and conception of the different test stands, the ICS team for their support and their reactivity and Yves Lussignol for the countless improvements and ideas he brought.